\documentclass[a4paper,12pt]{article}

\usepackage{jheppub}

\usepackage[T1]{fontenc}
\usepackage{lmodern}
\usepackage{amsfonts}
\usepackage{bm}
\usepackage{booktabs}
\usepackage{graphicx}
\usepackage{microtype}

\makeatletter
\gdef\@fpheader{Prepared for submission to Classical and Quantum Gravity}
\makeatother

\title{\boldmath Cauchy-horizon flux coefficients in the reduced Polyakov model}

\author[a,b]{Damien A.~Easson}

\affiliation[a]{Department of Physics, Arizona State University,\\
Tempe, Arizona 85287, USA}
\affiliation[b]{Beyond Center for Fundamental Concepts in Science,\\
Arizona State University, Tempe, Arizona 85287, USA}

\emailAdd{easson@asu.edu}

\abstract{
We study Cauchy-horizon flux amplification in the stationary reduced Polyakov
model for spherically symmetric charged black holes. The corresponding
four-dimensional coefficient depends on global scattering and on the quantum
state, while the reduced model gives an exactly calculable anomaly-induced
radial contribution. For a nonextremal inner horizon on a stationary background,
a finite late-time Eddington--Finkelstein flux is amplified by the affine
blueshift into a leading $V_-^{-2}$ divergence. In the Polyakov model the
coefficient is
$F_-^{(\infty)}=t_v-N\kappa_-^2/(48\pi)$, so the leading term vanishes only
when
$t_v=N\kappa_-^2/(48\pi)$. This inner-horizon condition is distinct from the
corresponding outer-horizon condition on $t_u$, while standard Unruh-type and
outer-horizon thermal choices generically lie away from it. Other quantum
contributions and classical tails enter through the total late-time flux.
Cancellation of the $V_-^{-2}$ term requires the total constant coefficient
$F_0$ to vanish, while asymptotically flat power-law tails produce
logarithmically weakened divergences and exponentially decaying de~Sitter
tails produce a different power of $V_-$. These results identify the
conditions controlling the leading Cauchy-horizon divergence and its
cancellation.
}

\begin{document}
\maketitle
\flushbottom

\section{Introduction}
\label{sec:intro}

Cauchy horizons are ubiquitous in charged, rotating, and NUT-charged electrovacuum solutions, with or without a cosmological constant, including the Reissner–Nordström and Kerr families and Taub–NUT geometries
\cite{Penrose:1964wq,Cardoso:2018nvb,Luna:2019olw,Holzegel:2006gn}. In standard
static or Boyer--Lindquist-type charts, the charged and rotating examples appear
as inner Killing horizons; their identification as Cauchy horizons is the
coordinate-independent statement that they bound a domain of dependence.
Their presence signals a
failure of global hyperbolicity in the unperturbed solution. The strong cosmic
censorship conjecture asserts that, for generic initial data, the maximal
Cauchy development is inextendible as a sufficiently regular Lorentzian
manifold. Classically, perturbations incident on the inner horizons of
Reissner--Nordstr\"om or Kerr black holes are infinitely blueshifted and give rise to mass inflation
and null singularity formation
\cite{Poisson:1990eh,Ori:1991zz,Brady:1995ni,Dafermos:2003wr}.
The compact Cauchy horizons of Taub--NUT have a different global structure and
require a separate treatment \cite{Holzegel:2006gn}; we return to this case in
section~\ref{sec:scope}.

Semiclassical physics adds a second source of inner-horizon stress. Quantum
fields in a black-hole background carry a renormalized stress-energy tensor
(RSET) whose
near-horizon behavior depends both on local geometry and on the global quantum
state. Four-dimensional analyses have shown complementary aspects of this
behavior. For Reissner--Nordstr\"om--de~Sitter black holes, Hollands, Wald, and
Zahn found a
leading \(V^{-2}\)-type divergence whose coefficient is independent of the
initial Hadamard state and generically nonzero
\cite{Markovic:1994gy,Hollands:2019whz,Hollands:2020qpe}. In asymptotically flat and
de~Sitter charged black holes, mode-sum and anomaly-induced calculations
determine corresponding finite horizon coefficients for specific quantum states
and scattering problems
\cite{Zilberman:2019buh,Zilberman:2022aum,Arrechea:2024ajt}. In the full
four-dimensional problem these coefficients depend on global scattering and on
the quantum state; stress-tensor conservation and anomaly information alone do
not fix them \cite{Birrell:1982ix}. Here we compute the stationary reduced
Polyakov contribution, including the leading Cauchy-horizon coefficient, its
cancellation condition, and the distinction between constant and tail-induced
divergences.

Our setting is the spherically symmetric reduction of four-dimensional
Einstein--Maxwell theory coupled to conformal matter. The reduced metric
describes the radial \((t,r)\) geometry, the area radius becomes a dilaton, and
the one-loop radial conformal anomaly is encoded by the Polyakov effective
action
\cite{Polyakov:1981rd,Christensen:1977jc,Callan:1992rs,Fabbri:2005mw}.
Near a nonextremal horizon, the radial null geometry is two-dimensional and
conformal, and the Polyakov term gives the anomaly-controlled contribution to
the reduced stress tensor
\cite{Unruh:1976db,Iso:2006ut,Kaloper:2012hu}. In this reduced action, the
anomaly-induced contribution is the stress tensor obtained by varying the
Polyakov functional with the chosen chiral state parameters. The resulting
stationary model gives the Cauchy-horizon coefficient exactly and allows direct
comparison with the corresponding outer-horizon cancellation condition.

The central object is the finite Eddington--Finkelstein coefficient
\(F_-^{(\infty)}\) approaching the right future Cauchy horizon, defined in
\eqref{eq:Fminus_infty}. Unless otherwise stated, stress-tensor components
through section~\ref{sec:totalflux} are reduced two-dimensional components; the
relation to four-dimensional \(s\)-wave components is given in
section~\ref{sec:curvature}. For a nonextremal inner horizon, the regular
near-horizon coordinate relation \eqref{eq:Vminus_def} converts a nonzero finite value of
\(F_-^{(\infty)}\) into a leading \(V_-^{-2}\) flux. The local exponential
factor is fixed by surface gravity; the coefficient is set by the quantum
state.

In the stationary Polyakov model, \eqref{eq:Fminus_polyakov} gives
\(F_-^{(\infty)}=t_v-N\kappa_-^2/(48\pi)\), where \(N\) is the effective
central charge and \(t_v\) is the incoming chiral state parameter. The leading
Polyakov term therefore vanishes on the inner-horizon cancellation surface
\eqref{eq:inner_cancellation_surface}. The future event horizon imposes the
distinct outgoing condition \eqref{eq:event_cancellation_surface}. These two
conditions define different lines in the stationary \((t_u,t_v)\) plane. Their
intersection cancels the leading quadratic Polyakov coefficients at both the
future event horizon and the right future Cauchy horizon; full regularity also
depends on subleading and non-Polyakov terms.

This state-space picture also places standard outer-horizon choices in the
same \((t_u,t_v)\) plane. The asymptotically flat Unruh-type choice sets
\(t_v=0\) and fixes \(t_u\) by cancellation of the leading coefficient at the
future event horizon. The outer-horizon thermal/Kubo--Martin--Schwinger (KMS)
choice instead sets \(t_u=t_v=N\kappa_+^2/(48\pi)\). Both generically lie away
from the inner-horizon cancellation surface
\eqref{eq:inner_cancellation_surface}. For nonextremal
Reissner--Nordstr\"om black holes, both therefore give nonzero reduced Polyakov
inner-horizon coefficients.

At late times the total stress tensor contains a constant term and decaying corrections. In
the asymptotically flat case, if additional quantum terms and classical tails give
\(T_{vv}^{\rm tot}=F_0+Av^{-p}+o(v^{-p})\), cancellation of the pure quadratic
term is the constant-level condition \(F_0=0\). Decaying Price-tail
\cite{Price:1971fb,Price:1972pw} terms cannot cancel a nonzero constant
coefficient; when the constant term is absent, they instead produce
logarithmically weakened divergences. Exponentially decaying tails appropriate
to de~Sitter backgrounds are treated separately in
section~\ref{sec:totalflux}. The corresponding radial null-curvature diagnostic
is discussed in section~\ref{sec:curvature}.

Dynamical Polyakov studies typically evolve or perturb specific spherical geometries
and find rapid outward inner-horizon motion
\cite{Barcelo:2020mjw,Boyanov:2025otp}. The anomaly-induced calculation of
Ref.~\cite{Arrechea:2024ajt} is closest in spirit to the present work. Both
use reduced anomaly-controlled stress tensors to probe Cauchy-horizon behavior.
The stationary background allows the two cancellation conditions and their relation
to the total flux to be written explicitly.

The conformal-gauge Polyakov stress tensor and the exponential conversion from
Eddington--Finkelstein to regular near-horizon coordinates are standard
ingredients, and previous two-dimensional anomaly calculations and
four-dimensional mode analyses already establish closely related horizon-flux
behavior
\cite{Christensen:1977jc,Markovic:1994gy,Zilberman:2019buh,Hollands:2019whz,Hollands:2020qpe,Zilberman:2022aum,Arrechea:2024ajt}.
A related $1+1$-dimensional analysis by Ju\'arez-Aubry and Louko studies
detector response and local energy density for specific HHI and Unruh states
\cite{JuarezAubry:2021vge}. Here we keep \(t_u\) and \(t_v\) arbitrary and
obtain the explicit inner-horizon coefficient
$C_-=t_v/\kappa_-^2-N/(48\pi)$, the distinct outer- and inner-horizon
cancellation conditions and their tuned intersection, the positions of the
standard outer choices, and the effect of decaying tails. A full four-dimensional mode calculation gives
the corresponding coefficient.

\begin{figure}[t]
\centering
\IfFileExists{rncauchy.png}
{\includegraphics[width=0.9\columnwidth]{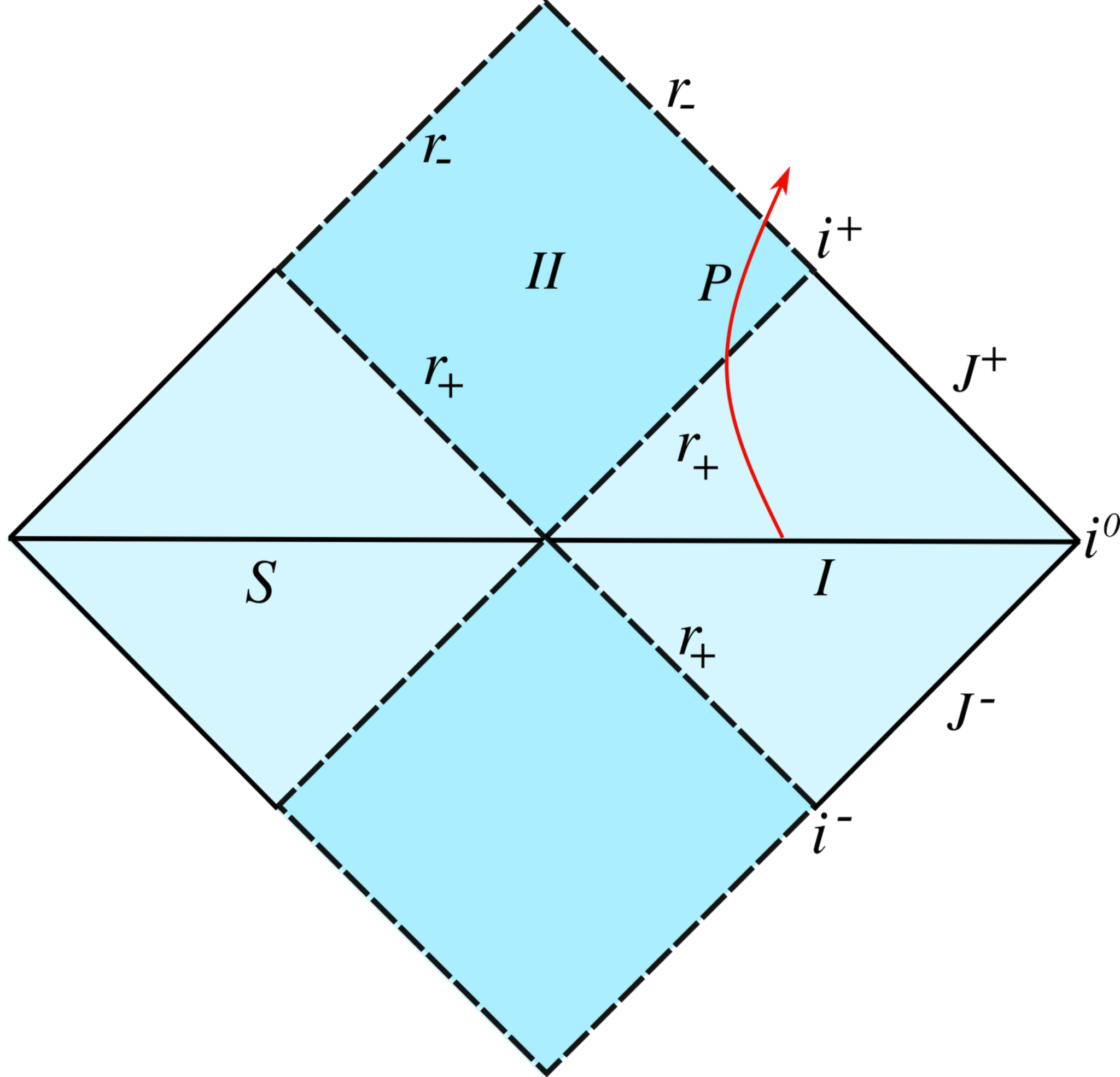}}
{\fbox{\parbox[c][1.9in][c]{0.86\columnwidth}{\centering Schematic causal diagram placeholder\\
(source image not included in this bundle).}}}
\caption{Causal structure of the Reissner--Nordstr\"om black-hole interior.
A timelike curve from an initial spacelike hypersurface crosses the event horizon
\(r_+\) and reaches the classical Cauchy horizon \(r_-\). In the reduced
fixed-background calculation studied here, the coordinate \(V_-\) regularizes
the right future Cauchy horizon, and a nonzero coefficient \(C_-\) produces a
radial flux component \(\langle T_{V_-V_-}\rangle\sim C_-/V_-^2\).\label{fig:CHdiagram}}
\end{figure}

This paper is organized as follows. In section~\ref{sec:affine} we derive the
affine amplification lemma, and in section~\ref{sec:framework} we introduce the
reduced model and stationary chiral parameters. In
sections~\ref{sec:coefficients} and \ref{sec:states} we compute the
inner-horizon coefficient and evaluate standard outer Polyakov choices.
In section~\ref{sec:totalflux} we analyze total fluxes and asymptotically flat
and de~Sitter tails; in section~\ref{sec:curvature} we give the conditional
radial curvature interpretation; and in section~\ref{sec:scope} we discuss the
scope of the reduced model. In appendix~\ref{app:dilaton_swave} we provide a
local dilaton-scaling check, and in appendix~\ref{app:polyakov_summary} we
summarize the Polyakov model and conventions.

\section{Cauchy-horizon boost}
\label{sec:affine}

The leading Cauchy-horizon behavior separates into a local kinematic part and a
state-dependent coefficient. The local part only uses the nonextremal
near-horizon relation between Eddington--Finkelstein time and a regular
Kruskal-like null coordinate that is asymptotically proportional to an affine
parameter along the crossing null direction.

Let \(v\) be the ingoing Eddington--Finkelstein coordinate approaching the
right future Cauchy horizon, and let \(V_-\) be such a regular Kruskal-like null
coordinate. For a nonextremal inner horizon,
\begin{equation}
V_-=-e^{-\kappa_- v},
\qquad
\kappa_->0,
\qquad
v\to+\infty .
\label{eq:Vminus_def}
\end{equation}
The overall normalization of \(V_-\) is immaterial. Equation
\eqref{eq:Vminus_def} gives
\begin{equation}
\frac{dV_-}{dv}=-\kappa_- V_-,
\qquad
\frac{dv}{dV_-}=-\frac{1}{\kappa_-V_-}.
\label{eq:affine_derivative}
\end{equation}
For a fixed physical state and an already-renormalized stress tensor, changing
from \(v\) to the regular coordinate \(V_-\) is an ordinary tensorial
transformation of components. The state and the tensor are held fixed; no new
normal-ordering prescription is defined in the \(V_-\) chart. Any Schwarzian
terms enter in the computation of \(\langle T_{vv}\rangle\), not in the
subsequent component transformation:
\begin{equation}
\langle T_{V_-V_-}\rangle
=
\left(\frac{dv}{dV_-}\right)^2
\langle T_{vv}\rangle
=
\frac{\langle T_{vv}\rangle}{\kappa_-^2V_-^2}.
\label{eq:tensorial_boost}
\end{equation}

\paragraph{Lemma 1: affine Cauchy-horizon amplification.}
If the ingoing Eddington--Finkelstein component has a finite late-time limit
\begin{equation}
F_-^{(\infty)}
\equiv
\lim_{v\to+\infty}\langle T_{vv}\rangle ,
\label{eq:Fminus_infty}
\end{equation}
then the Cauchy-horizon flux has pure quadratic coefficient
\begin{equation}
C_-
\equiv
\lim_{V_-\to0}V_-^2\langle T_{V_-V_-}\rangle
=
\frac{F_-^{(\infty)}}{\kappa_-^2}.
\label{eq:Cminus_general}
\end{equation}
Equivalently, when \(F_-^{(\infty)}\neq0\),
\begin{equation}
\langle T_{V_-V_-}\rangle
\sim
\frac{F_-^{(\infty)}}{\kappa_-^2V_-^2}
=
\frac{C_-}{V_-^2}.
\label{eq:affine_amp_result}
\end{equation}
If \(F_-^{(\infty)}=0\), the pure \(V_-^{-2}\) coefficient is absent and the
leading behavior is determined by subleading late-time terms in
\(\langle T_{vv}\rangle\).

\paragraph{Proof.}
Multiplying \eqref{eq:tensorial_boost} by \(V_-^2\) gives
\begin{equation}
V_-^2\langle T_{V_-V_-}\rangle
=
\frac{\langle T_{vv}\rangle}{\kappa_-^2}.
\end{equation}
Taking \(v\to+\infty\), equivalently \(V_-\to0\), gives
\eqref{eq:Cminus_general}. \(\square\)

\medskip
Near the Cauchy horizon, the regular Kruskal-like coordinate \(V_-\) is
asymptotically proportional, up to a finite nonzero factor, to an affine
parameter along the corresponding null direction. If \(\lambda\) is such an
affine parameter, then
\begin{equation}
V_-=\alpha(\lambda_0-\lambda)
+O\!\left((\lambda_0-\lambda)^2\right),
\qquad
\alpha\neq0 .
\label{eq:V_affine_lambda}
\end{equation}
Thus the divergence in \eqref{eq:affine_amp_result}, when \(C_-\neq0\), is
reached at a finite affine parameter \(\lambda\). The sign and magnitude of \(C_-\) are determined by the quantum state and by
any scattering or boundary conditions entering the late-time coefficient
\(F_-^{(\infty)}\), while the local boost fixes the amplification law.

The derivation above assumes a nonextremal inner horizon, \(\kappa_->0\). At extremality the inner horizon is
degenerate, the lapse has a multiple rather than a simple zero, and the regular
near-horizon coordinate is not related to \(v\) by
\eqref{eq:Vminus_def}. Consequently, the factors of \(\kappa_-^{-2}\) in the
nonextremal formulas cannot be continued by simply taking
\(\kappa_-\to0\); the near-horizon and extremal limits do not commute, and the
extremal case requires a separate analysis \cite{Trivedi:1992vh}.

\section{Reduced Polyakov framework}
\label{sec:framework}

The reduced model follows the fixed-charge spherical reduction conventions used
for four-dimensional Reissner--Nordstr\"om--de~Sitter black holes. Starting from
the Einstein--Maxwell theory with a cosmological constant
\cite{Grumiller:2002nm},
\begin{align}
S_4
={}&
\frac{1}{16\pi G_4}
\int d^4x\,\sqrt{-g^{(4)}}\,
\left(R^{(4)}-2\Lambda\right)
\nonumber\\
&-
\frac{1}{16\pi}
\int d^4x\,\sqrt{-g^{(4)}}\,
F_{\mu\nu}F^{\mu\nu} ,
\label{eq:S4_EML}
\end{align}
we assume a spherically symmetric geometry,
\begin{equation}
ds_4^2
=
g_{ab}(x)dx^adx^b+r^2(x)d\Omega_2^2,
\qquad
X=r^2.
\label{eq:spherical_ansatz}
\end{equation}
The squared area radius \(X\) is the two-dimensional dilaton. After integrating
over the two-sphere and eliminating the two-dimensional Maxwell field at a fixed
charge \(Q\), the reduced fixed-charge action is
\begin{equation}
S_2
=
\frac{1}{4G_4}
\int d^2x\,\sqrt{-g}\,
\left[
X R^{(2)}
+\frac{1}{2X}(\nabla X)^2
+2V(X)
\right],
\label{eq:S2_fixed_charge}
\end{equation}
with
\begin{equation}
V(X)
=
1-\Lambda X-\frac{G_4Q^2}{X}.
\label{eq:V_fixed_charge}
\end{equation}
The sign of the charge term in \eqref{eq:V_fixed_charge} is the fixed-charge
sign: after solving the Maxwell equation we work with the fixed-charge
Routhian rather than the naive on-shell Maxwell Lagrangian. We now show that
this convention reproduces the familiar four-dimensional
Reissner--Nordstr\"om--de~Sitter lapse.

Defining the radial primitive
\begin{equation}
W'(r)=V(r^2),
\label{eq:Wprime_def}
\end{equation}
one obtains
\begin{equation}
W(r)
=
r-\frac{\Lambda r^3}{3}+\frac{G_4Q^2}{r},
\label{eq:W_r}
\end{equation}
and the classical static solution of the fixed-charge reduced equations may be
written as
\begin{equation}
r\,\xi(r)=W(r)-2G_4M.
\label{eq:r_xi_W}
\end{equation}
Therefore
\begin{equation}
\xi(r;M,Q)
=
1-\frac{2G_4M}{r}
+\frac{G_4Q^2}{r^2}
-\frac{\Lambda r^2}{3}.
\label{eq:RNdS_lapse}
\end{equation}
In the asymptotically flat case used for the basic Reissner--Nordstr\"om
black-hole discussion, we have \(\Lambda=0\), and thus
\begin{equation}
\xi(r;M,Q)
=
1-\frac{2G_4M}{r}
+\frac{G_4Q^2}{r^2}.
\label{eq:RN_lapse_new}
\end{equation}

The coefficient calculation below requires only a fixed nonextremal static
background with an event horizon and a Cauchy horizon. Let \(r_+\) and \(r_-\)
denote the outer black-hole and inner horizons,
\begin{equation}
\xi(r_+)=0,
\qquad
\xi(r_-)=0,
\qquad
r_+>r_-,
\label{eq:horizon_roots_new}
\end{equation}
with positive surface-gravity magnitudes
\begin{equation}
\kappa_\pm=\frac12|\xi'(r_\pm)|.
\label{eq:kappa_pm_def_new}
\end{equation}
The presence of a cosmological horizon changes the global boundary conditions
that select \(t_u\) and \(t_v\). Once these parameters are specified, the local
calculation below depends only on the nonextremal near-horizon form of
\(\xi(r)\), so the derivation of the reduced Polyakov offset is unchanged. The
same fixed-charge reduced framework underlies
recent analyses of two-horizon evaporation and flux balance in
Schwarzschild--de~Sitter and Reissner--Nordstr\"om--de~Sitter black
holes~\cite{Easson:2025ekn,Easson:2026kng}.

The anomaly-induced contribution of the two-dimensional conformal fields is the
Polyakov effective action,
\begin{equation}
S_{\rm P}
=
-\frac{N}{96\pi}
\int d^2x\,\sqrt{-g}\,
R^{(2)}\Box^{-1}R^{(2)} ,
\label{eq:polyakov_action}
\end{equation}
where \(N\) is the effective central charge. In the conventions used here its
variation gives
\begin{equation}
\langle T^a{}_{a}\rangle
=
\frac{N}{24\pi}R^{(2)} .
\label{eq:trace_anomaly}
\end{equation}
The Polyakov action is the universal nonlocal functional that reproduces this
two-dimensional trace anomaly
\cite{Davies:1976ei,Christensen:1977jc,Polyakov:1981rd,Callan:1992rs,Riegert:1984kt}.
A convenient local representation introduces an auxiliary field \(\psi\):
\begin{align}
S_{\rm P}
={}&
-\frac{N}{96\pi}
\int d^2x\,\sqrt{-g}\,
\left[
(\nabla\psi)^2+2\psi R^{(2)}
\right],
\nonumber\\
&\hspace{3cm}
\Box\psi=R^{(2)} .
\label{eq:polyakov_local}
\end{align}
The field \(\psi\) is an auxiliary representation of the nonlocal response: its
homogeneous part and boundary conditions encode the choice of inverse
d'Alembertian rather than an additional propagating matter degree of freedom
\cite{Birrell:1982ix,Fabbri:2005mw}.

\subsection{Chiral state parameters}
\label{subsec:chiral_state_data}

In conformal coordinates,
\begin{equation}
ds_2^2=-e^{2\rho(u,v)}du\,dv ,
\label{eq:conformal_metric_new}
\end{equation}
the reduced Polyakov stress tensor has the chiral components
\begin{equation}
\langle T_{vv}\rangle
=
-\frac{N}{12\pi}
\left[
(\partial_v\rho)^2-\partial_v^2\rho
\right]
+t_v(v),
\label{eq:polyakov_Tvv_new}
\end{equation}
and
\begin{equation}
\langle T_{uu}\rangle
=
-\frac{N}{12\pi}
\left[
(\partial_u\rho)^2-\partial_u^2\rho
\right]
+t_u(u).
\label{eq:polyakov_Tuu_new}
\end{equation}
The functions \(t_v(v)\) and \(t_u(u)\) parametrize the quantum state along the
two null directions
\cite{Christensen:1977jc,Birrell:1982ix,Callan:1992rs,Grumiller:2002nm}. In
the stationary family considered here they are constants, denoted \(t_v\) and
\(t_u\). With this convention, \(t_u\) and \(t_v\) have the same dimensions as
the reduced stress-tensor components.

We distinguish the physical renormalized stress tensor from the coordinate-dependent
normal-ordered chiral representative used to specify the state. We denote this
representative by \({\cal T}_{vv}\). Under a reparametrization
\(v\mapsto V(v)\), we have
\begin{equation}
{\cal T}_{vv}
=
\left(\frac{dV}{dv}\right)^2
{\cal T}_{VV}
-
\frac{N}{24\pi}\{V,v\},
\label{eq:schwarzian_law_new}
\end{equation}
where
\begin{equation}
\{V,v\}
=
\frac{V'''}{V'}
-
\frac32\left(\frac{V''}{V'}\right)^2
\label{eq:schwarzian_def_new}
\end{equation}
is the Schwarzian derivative.

Equation~\eqref{eq:schwarzian_law_new} determines how the chiral representation
of the Polyakov stress changes when the null chart is changed. It is distinct
from the tensorial affine-frame relation in section~\ref{sec:affine}. Once a
renormalized stress tensor and state are fixed, its Cauchy-horizon
component is computed by the ordinary component transformation
\eqref{eq:tensorial_boost}.

A Hadamard state has the standard local short-distance singularity structure.
Extension across a horizon requires the renormalized stress tensor to remain
finite in a regular frame there, so cancellation of the leading divergent
coefficient is necessary. It is not sufficient: subleading, dilaton-dependent,
or other state-dependent terms may remain singular. We therefore use
``cancellation'' below only for the displayed leading coefficient.

\subsection{Static null coordinates}
\label{subsec:static_null_coordinates}

For the fixed static background used below, the two-dimensional line element is
\begin{equation}
ds_2^2=-\xi(r)dt^2+\xi(r)^{-1}dr^2 .
\label{eq:static_line_element}
\end{equation}
After introducing the tortoise coordinate
\begin{equation}
\frac{dr_*}{dr}=\xi^{-1}(r),
\label{eq:tortoise_def_new}
\end{equation}
and the null coordinates
\begin{equation}
u=t-r_*,
\qquad
v=t+r_*,
\label{eq:uv_def_new}
\end{equation}
we have
\begin{equation}
ds_2^2=-\xi(r)\,du\,dv .
\label{eq:static_double_null_metric_new}
\end{equation}

A regular outgoing affine coordinate on the future event horizon is
\begin{equation}
U_+=-e^{-\kappa_+u},
\label{eq:Uplus_def_new}
\end{equation}
so the horizon is located at \(U_+=0\). The right future Cauchy horizon is
described by the regular coordinate \(V_-\) introduced in
\eqref{eq:Vminus_def}, with the horizon at \(V_-=0\).

The constants \(t_u\) and \(t_v\) are fixed by exterior boundary conditions in
the reduced model: for example, the absence of an incoming flux from
\(\mathcal I^-\) in the asymptotically flat Unruh-type construction, a
single-temperature thermal/KMS bath, or cosmological-horizon conditions in
de~Sitter settings
\cite{Unruh:1976db,Markovic:1994gy,Iso:2006ut,Kaloper:2012hu,Hollands:2019whz,Zilberman:2019buh,Hollands:2020qpe,Zilberman:2022aum}.

\section{Inner-horizon coefficient}
\label{sec:coefficients}

We now compute the late-time Eddington--Finkelstein coefficient
\(F_-^{(\infty)}\) in this setting. The result uses
only the static near-horizon form of the two-dimensional metric and the chiral
state parameter \(t_v\).

For the static double-null metric \eqref{eq:static_double_null_metric_new}
one may write locally in each connected region
\begin{equation}
ds_2^2=-\sigma e^{2\rho}du\,dv,
\qquad
e^{2\rho}=|\xi|,
\qquad
\sigma=\operatorname{sgn}\xi,
\qquad
\rho=\frac12\ln|\xi|.
\label{eq:rho_xi_abs}
\end{equation}
Since $\sigma$ is constant within each connected region, the chiral derivative
terms are determined entirely by derivatives of $\rho$. The absolute value keeps the conformal
factor positive in regions where \(\xi\) changes sign. Away from the zero of \(\xi\),
\(\partial_r\ln|\xi|=\xi'/\xi\), so the local near-horizon expressions entering
the chiral Polyakov tensor are unchanged. Since
\begin{equation}
r_*=\frac{v-u}{2},
\qquad
\frac{dr_*}{dr}=\xi^{-1},
\label{eq:rstar_uv_relation}
\end{equation}
we have
\begin{equation}
\partial_v r=\frac{\xi}{2},
\qquad
\partial_u r=-\frac{\xi}{2}.
\label{eq:partial_r_uv}
\end{equation}
Therefore
\begin{equation}
\partial_v\rho
=
\frac12\frac{\xi'}{\xi}\partial_v r
=
\frac14\xi',
\qquad
\partial_u\rho
=
-\frac14\xi',
\label{eq:rho_first_derivatives}
\end{equation}
and
\begin{equation}
\partial_v^2\rho
=
\partial_u^2\rho
=
\frac18\xi\xi''.
\label{eq:rho_second_derivatives}
\end{equation}
The factors of \(\xi\) from \(\partial_v r\) and \(\partial_u r\) cancel the
\(1/\xi\) derivative of \(\ln|\xi|\), so these expressions are insensitive to
the sign of \(\xi\) on either side of a simple horizon.

Substituting \eqref{eq:rho_first_derivatives} and
\eqref{eq:rho_second_derivatives} into the stress tensor gives
\begin{equation}
\langle T_{vv}\rangle
=
-\frac{N}{192\pi}
\left(
\xi'^2-2\xi\xi''
\right)
+t_v,
\label{eq:Tvv_static_polyakov}
\end{equation}
and similarly
\begin{equation}
\langle T_{uu}\rangle
=
-\frac{N}{192\pi}
\left(
\xi'^2-2\xi\xi''
\right)
+t_u .
\label{eq:Tuu_static_polyakov}
\end{equation}
At a nonextremal horizon \(r=r_h\),
\begin{equation}
\xi(r_h)=0,
\qquad
\kappa_h=\frac12|\xi'(r_h)|,
\label{eq:kappa_h_coeff}
\end{equation}
so
\begin{equation}
\lim_{r\to r_h}\langle T_{vv}\rangle
=
t_v-\frac{N}{48\pi}\kappa_h^2,
\qquad
\lim_{r\to r_h}\langle T_{uu}\rangle
=
t_u-\frac{N}{48\pi}\kappa_h^2.
\label{eq:horizon_limits_chiral}
\end{equation}

\paragraph{Lemma 2: stationary Polyakov horizon limit.}
At the right future Cauchy horizon, the late-time
Eddington--Finkelstein flux is
\begin{equation}
F_-^{(\infty)}
=
\lim_{v\to+\infty}\langle T_{vv}\rangle
=
t_v-\frac{N}{48\pi}\kappa_-^2 .
\label{eq:Fminus_polyakov}
\end{equation}

\paragraph{Proof.}
Along the right future Cauchy horizon, \(v\to+\infty\) and \(r\to r_-\).
Applying \eqref{eq:horizon_limits_chiral} with \(r_h=r_-\) gives
\eqref{eq:Fminus_polyakov}. \(\square\)

Combining Lemma~1 with Lemma~2 gives the inner-horizon coefficient.

\paragraph{Proposition 1: inner-horizon coefficient.}
At the right future Cauchy horizon, the pure quadratic coefficient is
\begin{equation}
C_-
=
\frac{1}{\kappa_-^2}
\left(
t_v-\frac{N}{48\pi}\kappa_-^2
\right)
=
\frac{t_v}{\kappa_-^2}
-\frac{N}{48\pi}.
\label{eq:Cminus_polyakov}
\end{equation}
For \(C_-\neq0\),
\begin{equation}
\langle T_{V_-V_-}\rangle
\sim
\frac{C_-}{V_-^2}.
\label{eq:leading_affine_polyakov}
\end{equation}
The \(V_-^{-2}\) term vanishes precisely when
\begin{equation}
t_v=\frac{N}{48\pi}\kappa_-^2 .
\label{eq:inner_cancellation_surface}
\end{equation}

\paragraph{Proof.}
Lemma~1 gives \(C_-=F_-^{(\infty)}/\kappa_-^2\). Substituting
\eqref{eq:Fminus_polyakov} gives \eqref{eq:Cminus_polyakov}. Setting
\(C_-=0\) gives \eqref{eq:inner_cancellation_surface}. \(\square\)

\medskip
Equation~\eqref{eq:inner_cancellation_surface} controls the pure
\(V_-^{-2}\) Polyakov coefficient. Full semiclassical regularity depends on
subleading terms and on any further contributions to the total renormalized
stress tensor.

\section{State-space picture and standard outer Polyakov choices}
\label{sec:states}

Our model is characterized by the two constants
\((t_u,t_v)\). Cancellation of the leading term at the future event
horizon fixes the outgoing parameter \(t_u\), while the corresponding
cancellation at the right future Cauchy horizon fixes the incoming parameter
\(t_v\).

\subsection{Event-horizon cancellation surface}
\label{subsec:event_surface}

Near the future event horizon, a regular outgoing affine coordinate is
\(U_+=-e^{-\kappa_+u} \).
Applying the tensorial argument of \eqref{eq:tensorial_boost} with
\((v,V_-,\kappa_-)\) replaced by \((u,U_+,\kappa_+)\), and using the horizon
limit \eqref{eq:horizon_limits_chiral} at \(r=r_+\), the
pure quadratic coefficient at the future event horizon is
\begin{equation}
C_+
=
\frac{1}{\kappa_+^2}
\left(
t_u-\frac{N}{48\pi}\kappa_+^2
\right).
\label{eq:Cplus_polyakov}
\end{equation}
Thus the leading Polyakov term at the future event horizon is absent
precisely when
\begin{equation}
t_u=\frac{N}{48\pi}\kappa_+^2.
\label{eq:event_cancellation_surface}
\end{equation}
Together with the inner-horizon condition
\eqref{eq:inner_cancellation_surface}, this defines two distinct cancellation
surfaces. Their intersection is
\begin{equation}
(t_u,t_v)_*
=
\left(
\frac{N}{48\pi}\kappa_+^2,
\frac{N}{48\pi}\kappa_-^2
\right).
\label{eq:tuned_intersection}
\end{equation}
This point cancels the leading quadratic Polyakov coefficients at both the
future event horizon and the right future Cauchy horizon; its physical status
is discussed below.

\subsection{Physical status of the tuned intersection}
\label{subsec:tuned_state_status}

The \((t_u,t_v)\) plane parametrizes stationary chiral choices for the
conserved reduced Polyakov stress tensor on the fixed static region. Extending
such a choice to a Hadamard state regular on the relevant horizon branches
imposes additional global conditions. In particular, the intersection
\eqref{eq:tuned_intersection} imposes only the two leading-coefficient
cancellations used to define it.

For an asymptotically flat exterior, the geometric terms in
\eqref{eq:polyakov_Tvv_new} and \eqref{eq:polyakov_Tuu_new} vanish at infinity,
so the asymptotic incoming and outgoing chiral fluxes are \(t_v\) and \(t_u\),
respectively. The tuned point therefore carries the unequal constant fluxes
\(N\kappa_-^2/(48\pi)\) and \(N\kappa_+^2/(48\pi)\). For a nonextremal
Reissner--Nordstr\"om black hole it is neither empty at past null infinity nor a
single-temperature KMS state. The tuned point may be interpreted locally as a fine-tuned two-temperature chiral configuration; its global realization would require compatible boundary conditions.

Cancellation on
the past event-horizon branch would require
\(t_v=N\kappa_+^2/(48\pi)\), whereas cancellation on the opposite future
Cauchy-horizon branch would require
\(t_u=N\kappa_-^2/(48\pi)\). These conditions are incompatible with
\eqref{eq:tuned_intersection} when \(\kappa_+\ne\kappa_-\). Moreover, a
constant incoming flux maintained for arbitrarily long advanced time is not the
state produced by ordinary collapse with finite incoming energy. Thus the tuned intersection identifies simultaneous leading cancellation on
the two selected future branches, rather than a generic globally horizon-regular
state.

\subsection{Unruh-type and outer-horizon thermal/KMS Polyakov choices}
\label{subsec:outer_prescriptions}

Here ``Unruh-type'' and ``outer-KMS'' refer to the corresponding stationary
choices of \(t_u\) and \(t_v\). Four-dimensional coefficients also
include global scattering and additional field content, represented in
section~\ref{sec:totalflux} by \(F_{\rm other}^{(\infty)}\). The Unruh-type
choice is the fixed-background stationary analogue of the usual Unruh state,
while ``outer-KMS'' denotes thermal chiral parameters at the outer-horizon
temperature. When the horizon surface gravities differ, no single-temperature
KMS state is regular on every horizon branch
\cite{Birrell:1982ix,Markovic:1994gy}.

The asymptotically flat Unruh-type Polyakov choice sets the incoming chiral
parameter to zero,
\begin{equation}
t_v=0,
\label{eq:unruh_tv}
\end{equation}
and fixes the outgoing parameter by canceling the leading term at the future
event horizon \eqref{eq:event_cancellation_surface}.
Using \eqref{eq:Cminus_polyakov}, the right future Cauchy-horizon coefficient is
\begin{equation}
C_-^{\rm Unruh}
=
-\frac{N}{48\pi}.
\label{eq:Cminus_unruh}
\end{equation}

Formal outer-horizon thermal/KMS Polyakov choices at temperature \(T\) have
equal stationary chiral parameters
\begin{equation}
t_u=t_v=\frac{\pi N}{12}T^2.
\label{eq:thermal_state_data}
\end{equation}
For the outer-horizon temperature
\begin{equation}
T_+=\frac{\kappa_+}{2\pi},
\label{eq:outer_hawking_temperature}
\end{equation}
this gives
\begin{equation}
t_u=t_v=\frac{N}{48\pi}\kappa_+^2,
\label{eq:outer_kms_data}
\end{equation}
and the corresponding inner-horizon coefficient is
\begin{equation}
C_-^{\rm outer\mbox{-}KMS}
=
\frac{N}{48\pi}
\left(
\frac{\kappa_+^2}{\kappa_-^2}-1
\right).
\label{eq:Cminus_outer_kms}
\end{equation}

For the asymptotically flat Reissner--Nordstr\"om family in the fixed-charge
conventions of section~\ref{sec:framework}, $\xi$ is given by \eqref{eq:RN_lapse_new}
with
\begin{equation}
r_\pm
=
G_4M
\pm
\sqrt{G_4^2M^2-G_4Q^2}.
\label{eq:RN_horizons_states}
\end{equation}
The positive surface-gravity magnitudes are
\begin{equation}
\kappa_+
=
\frac{r_+-r_-}{2r_+^2},
\qquad
\kappa_-
=
\frac{r_+-r_-}{2r_-^2}.
\label{eq:RN_kappas_states}
\end{equation}
Hence
\begin{equation}
\frac{\kappa_+^2}{\kappa_-^2}
=
\frac{r_-^4}{r_+^4}
<1,
\label{eq:kappa_ratio_RN}
\end{equation}
and
\begin{equation}
C_-^{\rm outer\mbox{-}KMS}<0.
\label{eq:Cminus_kms_negative}
\end{equation}

\paragraph{Proposition 2: standard outer Polyakov choices.}
The asymptotically flat Unruh-type and outer-horizon thermal/KMS choices both
give nonzero pure quadratic coefficients at the right future Cauchy horizon,
given by \eqref{eq:Cminus_unruh} and \eqref{eq:Cminus_outer_kms}. For
nonextremal asymptotically flat Reissner--Nordstr\"om black holes,
\(C_-^{\rm outer\mbox{-}KMS}<0\).

\paragraph{Proof.}
The Unruh-type Polyakov choice gives \eqref{eq:unruh_tv} and
\eqref{eq:event_cancellation_surface}; substituting \(t_v=0\) into
\eqref{eq:Cminus_polyakov} gives \eqref{eq:Cminus_unruh}. The thermal/KMS
relation \eqref{eq:thermal_state_data}, evaluated at
\(T_+=\kappa_+/(2\pi)\), gives \eqref{eq:outer_kms_data}; substituting this
value of \(t_v\) into \eqref{eq:Cminus_polyakov} gives
\eqref{eq:Cminus_outer_kms}. Finally, \eqref{eq:RN_kappas_states} gives
\eqref{eq:kappa_ratio_RN}, so the outer-KMS coefficient is negative for
nonextremal asymptotically flat Reissner--Nordstr\"om black holes. \(\square\)

\begin{figure}[t]
\centering
\includegraphics[width=\columnwidth]{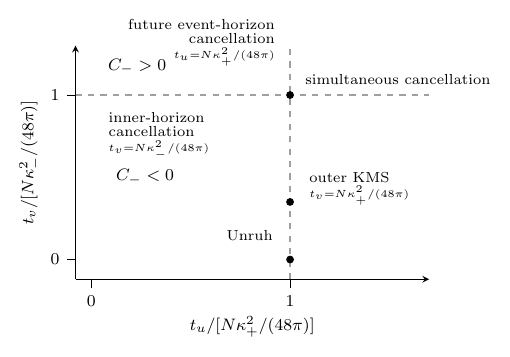}
\caption{Stationary \((t_u,t_v)\) parameter space for the reduced Polyakov model.
The dashed vertical and horizontal lines denote the future event-horizon and
inner-horizon cancellation surfaces, respectively, with their intersection
marking simultaneous cancellation of the leading coefficients. Since
the vertical coordinate is normalized by \(N\kappa_-^2/(48\pi)\), the
outer-KMS point lies below the inner-horizon cancellation line for
nonextremal asymptotically flat Reissner--Nordstr\"om black holes, where
\(\kappa_+<\kappa_-\).\label{fig:state_space}}
\end{figure}

\medskip
Whether or not the reduced Polyakov coefficient vanishes is distinct from the sign it
has when nonzero. A negative coefficient means that this radial Polyakov
component violates the null energy condition and gives defocusing, rather than
the standard positive-energy focusing, in the formal null-contracted Einstein
diagnostic. Its magnitude nevertheless has the same \(V_-^{-2}\) amplification
and marks breakdown of the fixed-background test-field approximation. The sign
of the full RSET coefficient can differ from the Polyakov contribution and is
the quantity relevant to focusing, as discussed in section~\ref{sec:curvature}. The stationary state-space picture is summarized in 
Figure~\ref{fig:state_space}.

\section{Total fluxes and asymptotic cancellation}
\label{sec:totalflux}

The Polyakov coefficient computed above is one contribution to the
semiclassical stress tensor. On the fixed background, the relevant quantity is
the total late-time \(T_{vv}\) coefficient, including finite quantum terms
outside the reduced Polyakov model as well as decaying classical tails. These contributions enter at different asymptotic orders, so cancellation
of the leading constant coefficient must be distinguished from the weaker
divergences generated by late-time tails.

\subsection{Asymptotically flat power-law tails}
\label{subsec:power_tails}

For asymptotically flat black holes, Price-law behavior motivates the inverse-power
expansion \cite{Price:1971fb,Price:1972pw}
\begin{equation}
T_{vv}^{\rm tot}(v)
=
F_0
+
A v^{-p}
+
o(v^{-p}),
\qquad
p>0.
\label{eq:total_flux_expansion}
\end{equation}
Here \(F_0\) denotes the total constant late-time coefficient,
\begin{equation}
F_0
=
F_{\rm P}^{(\infty)}
+
F_{\rm other}^{(\infty)} ,
\label{eq:F0_decomposition}
\end{equation}
where
\begin{equation}
F_{\rm P}^{(\infty)}
=
t_v-\frac{N}{48\pi}\kappa_-^2
\label{eq:FP_infty_totalflux}
\end{equation}
is the reduced Polyakov contribution and \(F_{\rm other}^{(\infty)}\) denotes
any additional finite contribution to the renormalized stress tensor at the
same asymptotic order. The term \(A v^{-p}\) represents a decaying Price-tail
contribution.

Allowing for an arbitrary normalization of \(V_-\),
\begin{equation}
V_-=-V_0e^{-\kappa_-v},
\qquad
v=\frac{1}{\kappa_-}\ln\frac{V_0}{|V_-|}.
\label{eq:v_log_relation}
\end{equation}
The constant \(V_0\) only changes subleading logarithmic terms, so it will be
set to unity in the power-law formulas below. We have
\begin{equation}
v^{-p}
=
\frac{\kappa_-^p}
{[\ln(1/|V_-|)]^p}
\left[1+o(1)\right].
\label{eq:v_tail_log}
\end{equation}
The tensorial transformation \eqref{eq:tensorial_boost} gives
\begin{equation}
T_{V_-V_-}^{\rm tot}
=
\frac{F_0}{\kappa_-^2V_-^2}
+
\frac{A\kappa_-^{p-2}}
{V_-^2[\ln(1/|V_-|)]^p}
+\cdots .
\label{eq:total_affine_expansion}
\end{equation}

\paragraph{Proposition 3: total-flux cancellation.}
For a late-time expansion of the form \eqref{eq:total_flux_expansion}, the
pure quadratic coefficient is
\begin{equation}
C_-^{\rm tot}
=
\frac{F_0}{\kappa_-^2}.
\label{eq:Cminus_total}
\end{equation}
If \(F_0\neq0\), this term dominates the near-horizon expansion: the first term
in \eqref{eq:total_affine_expansion} is then the leading one.
Cancellation of the pure \(V_-^{-2}\) coefficient is the constant-level
condition
\begin{equation}
F_0=0.
\label{eq:constant_level_cancellation}
\end{equation}
When \(A\neq0\), decaying tails still give logarithmically weakened
divergences,
\begin{equation}
T_{V_-V_-}^{\rm tot}
\sim
\frac{A\kappa_-^{p-2}}
{V_-^2[\ln(1/|V_-|)]^p}.
\label{eq:tail_affine_divergence}
\end{equation}

\paragraph{Proof.}
Substituting \eqref{eq:total_flux_expansion} into
\eqref{eq:tensorial_boost} and using \eqref{eq:v_tail_log} gives
\eqref{eq:total_affine_expansion}. The remaining statements follow by setting
\(F_0\) and \(A\) to zero or nonzero as indicated. \(\square\)

\medskip
For the Polyakov contribution alone, \(F_0=F_{\rm P}^{(\infty)}\) with
\(F_{\rm P}^{(\infty)}\) given by \eqref{eq:FP_infty_totalflux}, so
\eqref{eq:constant_level_cancellation} reduces to the cancellation surface
\eqref{eq:inner_cancellation_surface}. With further quantum contributions
included, the constant-level condition becomes
\begin{equation}
F_{\rm P}^{(\infty)}
+
F_{\rm other}^{(\infty)}
=0.
\label{eq:full_constant_cancellation}
\end{equation}
In this case a classical Price tail of the form \(A v^{-p}\) cannot cancel a nonzero
constant coefficient because its ratio to the constant term scales as
\begin{equation}
\frac{A v^{-p}}{F_0}
\to0
\qquad
(v\to+\infty),
\qquad
F_0\neq0.
\label{eq:tail_ratio}
\end{equation}
Its role is to determine the logarithmically subleading
structure, or the leading structure once the constant coefficient is tuned away.

The sign of \(F_0\) is relevant for focusing when the pure quadratic term is
present, and the discussion above concerns the presence or absence of the pure
quadratic coefficient. If \(F_0\neq0\), the sign of this total constant
coefficient determines the sign of the leading null-contracted curvature in the
Einstein equation. If \(F_0=0\), the strength of the remaining
tail-induced curvature divergence depends on the logarithmic power \(p\).

\subsection{Exponentially decaying de~Sitter tails}
\label{subsec:desitter_tails}

The late-time behavior depends on the asymptotic structure of the spacetime.
While \eqref{eq:total_flux_expansion} describes the inverse-power tails of the
asymptotically flat case, perturbations in asymptotically de~Sitter black-hole
spacetimes decay exponentially under the usual spectral assumptions
\cite{Hintz:2016gwb,Cardoso:2018nvb,Hollands:2019whz}. We therefore write
\begin{equation}
T_{vv}^{\rm tot}(v)
=
F_0+A e^{-\alpha v}+o(e^{-\alpha v}),
\qquad
\alpha>0.
\label{eq:desitter_flux_expansion}
\end{equation}
Here \(\alpha\) denotes the decay exponent of the flux component
\(T_{vv}^{\rm tot}\) itself. If a spectral gap is instead quoted for a field
amplitude, the corresponding stress-tensor decay exponent should be used in
\eqref{eq:desitter_flux_expansion}.
Restoring the normalization \(V_0\) in \eqref{eq:v_log_relation} and
defining
\begin{equation}
\beta=\frac{\alpha}{\kappa_-},
\qquad
e^{-\alpha v}=\left(\frac{|V_-|}{V_0}\right)^{\beta},
\label{eq:desitter_beta}
\end{equation}
the tensorial transformation gives
\begin{equation}
T_{V_-V_-}^{\rm tot}
=
\frac{F_0}{\kappa_-^2V_-^2}
+
\frac{A}{\kappa_-^2V_-^2}
\left(\frac{|V_-|}{V_0}\right)^{\beta}
+o\!\left(|V_-|^{\beta-2}\right).
\label{eq:desitter_tail_affine}
\end{equation}
Thus a nonzero \(F_0\) again gives the pure quadratic term. If \(F_0=0\), the
exponential tail diverges for \(0<\beta<2\), approaches a finite constant for
\(\beta=2\), and vanishes for \(\beta>2\). This affine-power classification is
distinct from the logarithmically weakened asymptotically flat behavior in
\eqref{eq:tail_affine_divergence}.

\section{Radial null-curvature diagnostic}
\label{sec:curvature}

We use the fixed-background RSET in a diagnostic extrapolation to infer the
radial curvature it would source through the semiclassical Einstein equation.
This is not a self-consistent semiclassical solution. This relation is conditional once
the stress becomes large enough for backreaction to matter. The component
\(T_{V_-V_-}\) is the stress tensor contracted with the radial null direction
associated with the regular coordinate \(V_-\). In the spherical \(s\)-wave normalization,
\begin{equation}
T_{V_-V_-}^{(2)}
=
4\pi r^2 T_{V_-V_-}^{(4)} .
\label{eq:swave_stress_relation}
\end{equation}
In a self-consistent spherical semiclassical geometry the null-contracted
Einstein equation would give
\begin{equation}
R^{(4)}_{V_-V_-}
=
8\pi G_4 T_{V_-V_-}^{(4)}
=
\frac{2G_4}{r^2}T_{V_-V_-}^{(2)} .
\label{eq:Rkk_swave_relation}
\end{equation}
The trace and cosmological-constant terms drop out after contraction with a
null vector.

Formally retaining the fixed-background limit \(r\to r_-\), a nonzero pure
quadratic coefficient would therefore source
\begin{equation}
R^{(4)}_{V_-V_-}
\sim
\frac{2G_4}{r_-^2}
\frac{C_-^{\rm tot}}{V_-^2}.
\label{eq:Rkk_Cminus}
\end{equation}
Using the affine relation \eqref{eq:V_affine_lambda}, the curvature contracted
with the corresponding affinely normalized tangent
\(k^a=dx^a/d\lambda\) is
\begin{equation}
R^{(4)}_{kk}
=
\left(\frac{dV_-}{d\lambda}\right)^2
R^{(4)}_{V_-V_-}
\sim
\frac{\mathcal K}{(\lambda_0-\lambda)^2},
\qquad
\mathcal K=
\frac{2G_4 C_-^{\rm tot}}{r_-^2}.
\label{eq:Rkk_lambda_scaling}
\end{equation}

\paragraph{Conditional radial null-curvature diagnostic.}
If \(C_-^{\rm tot}\neq0\) and the same null boundary and affine relation persist
into the backreacted regime, the fixed-background scaling would source a
divergent radial null Ricci component in a parallelly propagated frame:
\begin{equation}
|R^{(4)}_{kk}|
\sim
(\lambda_0-\lambda)^{-2}.
\label{eq:Rkk_pp_blowup}
\end{equation}
Once the RSET becomes large, the fixed-background approximation breaks down
and the subsequent geometry must be determined dynamically.

When the total coefficient has the focusing sign,
\begin{equation}
\mathcal K>0,
\label{eq:focusing_sign}
\end{equation}
the associated Tipler double integral diverges logarithmically \cite{Tipler:1977zza}:
\begin{equation}
\int^\lambda d\lambda'
\int^{\lambda'} d\lambda''
R^{(4)}_{kk}(\lambda'')
\sim
-\mathcal K\ln(\lambda_0-\lambda)
\to +\infty .
\label{eq:tipler_log_divergence}
\end{equation}
Thus, under the same conditional continuation, the pure quadratic term would
give Tipler-strength focusing when its total coefficient has the focusing sign.

If the pure quadratic coefficient is cancelled, \(F_0=0\), and \(A\neq0\),
the conditional diagnostic for an asymptotically flat background may instead
be controlled by a Price-tail term. From
\eqref{eq:tail_affine_divergence} we get
\begin{equation}
R^{(4)}_{kk}
\sim
\frac{\mathcal B}
{(\lambda_0-\lambda)^2
[\ln(1/(\lambda_0-\lambda))]^p},
\label{eq:Rkk_tail_only}
\end{equation}
with \(\mathcal B\) proportional to \(A\). For the focusing sign
\(\mathcal B>0\), the corresponding Tipler double integral has the asymptotic
form
\begin{align}
\int^\lambda d\lambda'
\int^{\lambda'} d\lambda''
R^{(4)}_{kk}(\lambda'')
&\sim
\mathcal B
\int^{\infty}
\frac{dL}{L^p},
\nonumber\\
L&=\ln\frac{1}{\lambda_0-\lambda}.
\label{eq:tail_tipler_integral}
\end{align}
It diverges for \(p\le1\) and converges for \(p>1\). Thus cancellation of the
constant coefficient changes the strength classification of the remaining
tail-induced curvature, even though the component itself still diverges
for any finite \(p>0\).

For an exponentially decaying de~Sitter tail with \(F_0=0\),
\eqref{eq:desitter_tail_affine} instead gives the conditional scaling
\begin{equation}
R^{(4)}_{kk}
\sim
\mathcal D(\lambda_0-\lambda)^{\beta-2},
\qquad
\beta=\frac{\alpha}{\kappa_-}>0,
\label{eq:Rkk_desitter_tail}
\end{equation}
with \(\mathcal D\) proportional to \(A\). It diverges only for
\(0<\beta<2\), and its Tipler double integral converges for every
\(\beta>0\). The exponential-tail strength classification is therefore weaker
than the power-law classification in \eqref{eq:tail_tipler_integral}.

For the Polyakov contribution alone, the Unruh-type and outer-horizon
thermal/KMS Polyakov choices analyzed in section~\ref{sec:states} give negative
\(C_-\) for nonextremal asymptotically flat Reissner--Nordstr\"om black holes. Those
examples demonstrate the pure \(V_-^{-2}\) divergence of the radial
Polyakov component and, by themselves, have the defocusing sign. The focusing
interpretation belongs to the sign of the total coefficient
\(C_-^{\rm tot}\).

\subsection{Fixed Cauchy horizon and dynamical marginal surfaces}
\label{subsec:dynamical_horizons}

The fixed-background limit above should be separated from a dynamical
trapping-horizon analysis. In spherical double-null coordinates one may write
\begin{equation}
ds^2=-\Omega^2(u,v)\,du\,dv+r^2(u,v)d\Omega_2^2,
\qquad
\theta_{(u)}=\frac{2\,\partial_u r}{r},
\qquad
\theta_{(v)}=\frac{2\,\partial_v r}{r}.
\label{eq:double_null_expansions}
\end{equation}
The zero sets \(\partial_u r=0\) and \(\partial_v r=0\) define the two families
of radial marginal surfaces. Their association with the ingoing and outgoing
expansions depends on the orientation of the null coordinates and the
double-null convention. In a stationary solution an inner marginal surface
may coincide with an inner Killing horizon and with the relevant Cauchy-horizon
branch. In a dynamical geometry these are distinct notions: the two marginal
surfaces can move or change causal character, while a Cauchy horizon is defined
globally as a boundary of a domain of dependence.

Throughout this analysis, \(r\) and \(\Omega\) are held fixed at their
static-background values. Classical and semiclassical double-null studies show
that the inner apparent horizon can move, become spacelike, or separate from
the Cauchy horizon, and Polyakov backreaction can drive rapid outward
inner-horizon motion
\cite{Oren:2003qc,Hong:2008mw,Barcelo:2020mjw,Chew:2023jti,Boyanov:2025otp}.
Our result concerns flux amplification at the classical Cauchy horizon
and the point at which the test-field approximation fails. The subsequent
trajectories of the marginal surfaces and the existence or causal character of
any Cauchy horizon belong to the self-consistent double-null evolution.

\section{The reduced model}
\label{sec:scope}

The coefficient formula derived in section~\ref{sec:coefficients} is local at the
nonextremal Cauchy horizon and exact within the stationary Polyakov model. Its
inputs are the near-horizon relation \(V_-\sim e^{-\kappa_-v}\), the stationary
chiral parameter \(t_v\), and the Polyakov anomaly coefficient \(N\). 

In a full spherical reduction of four-dimensional matter, the area radius
\(r\) appears as a dilaton coupling. Such terms can change the subleading
near-horizon structure and introduce additional state-dependent contributions. For
the static near-horizon scaling used here, however, the local dilaton
derivative terms satisfy
\begin{equation}
\partial_v\Phi=O(\xi),
\qquad
\partial_v^2\Phi=O(\xi),
\qquad
e^{-2\Phi}=r^2 .
\label{eq:scope_dilaton_scaling}
\end{equation}
This local scaling shows that the derivative structures displayed explicitly in
appendix~\ref{app:dilaton_swave}, and their products with finite
\(\partial_v\rho\), do not supply a constant-order shift to the pure Polyakov
geometric offset \(-N\kappa_-^2/(48\pi)\). Finite nonlocal, homogeneous, or
additional state-dependent contributions are included in
\(F_{\rm other}^{(\infty)}\) in section~\ref{sec:totalflux}.

Additional quantum fields, greybody effects, and higher-dimensional scattering also
contribute to the total constant coefficient \(F_0\) in
section~\ref{sec:totalflux}. The split between
\(F_{\rm P}^{(\infty)}\) and \(F_{\rm other}^{(\infty)}\) separates the reduced
Polyakov contribution from the remaining terms. A further finite contribution at the same
asymptotic order can shift the pure quadratic coefficient,
\begin{equation}
C_-^{\rm tot}
=
\frac{F_{\rm P}^{(\infty)}+F_{\rm other}^{(\infty)}}{\kappa_-^2}.
\label{eq:scope_total_C}
\end{equation}
Determining \(F_{\rm other}^{(\infty)}\) requires a global calculation. Within a
dilaton-coupled two-dimensional model one must specify the complete nonlocal
effective action, its Green function or homogeneous solutions, and the global
state and boundary conditions, and then evaluate the full RSET. For the actual
four-dimensional \(s\)-wave coefficient one must instead solve the radial mode
scattering problem with the chosen state, include the greybody factors, perform
a renormalized mode-sum or equivalent point-splitting calculation, and take the
late-time inner-horizon limit. Decaying tails contribute either the
logarithmically weakened terms in \eqref{eq:total_affine_expansion} or the
de~Sitter affine powers in \eqref{eq:desitter_tail_affine}.

For rotating black holes, the local amplification remains the same once the
full theory supplies a finite coefficient \(F_-^{(\infty)}\). Computing that
coefficient in Kerr or Kerr--Newman is a four-dimensional mode problem, with
angular modes, superradiance, and non-axisymmetric contributions entering the
renormalized stress tensor \cite{Zilberman:2022aum}. Equation
\eqref{eq:Cminus_polyakov} gives the corresponding result in the spherical
radial model, where the anomaly-induced contribution is calculable. A
rotating cancellation condition follows from the full mode coefficient.

For Taub--NUT and NUT-charged black holes, the same local tensorial boost applies
near any nondegenerate Cauchy-horizon generator once a suitable finite
pre-affine flux component is known. Determining the coefficient requires the
compact or fibered horizon structure, angular modes, Misner-string or
global-identification issues, and the admissibility of global quantum states
\cite{Holzegel:2006gn,Rahman:2020vsv}. Thus the same local boost suggests the
form of the amplification, while the NUT coefficient and any cancellation
condition require a separate calculation.

\section{Discussion}
\label{sec:discussion}

The stationary reduced Polyakov model gives an exact analytic expression for
the anomaly-induced radial contribution to the Cauchy-horizon coefficient on a
fixed nonextremal background. The state dependence enters through the incoming
chiral parameter \(t_v\), while the local amplification is fixed by
\(V_-\sim e^{-\kappa_-v}\). The inner-horizon cancellation condition
\(t_v=N\kappa_-^2/(48\pi)\) is distinct from the future event-horizon condition
\(t_u=N\kappa_+^2/(48\pi)\). Standard outer Polyakov choices therefore
generally leave a nonzero leading inner-horizon term. The tuned intersection
cancels the leading terms on the two selected future branches, while global
regularity imposes additional conditions discussed in
section~\ref{subsec:tuned_state_status}.
The calculation separates the universal local blueshift from the
state-dependent coefficient that must ultimately be determined by the global
quantum problem. In this sense, the reduced result provides an analytic
reference point for comparison with four-dimensional mode calculations and
self-consistent semiclassical evolution.

For the total stress tensor, constant terms and decaying tails play different
roles. Removing the pure
\(V_-^{-2}\) term requires cancellation of the total constant
Eddington--Finkelstein flux, \(F_0=0\). Decaying Price tails cannot cancel a
nonzero constant coefficient, but when the constant term is absent they control
the remaining logarithmically weakened near-horizon behavior in asymptotically
flat backgrounds. Exponentially decaying de~Sitter tails give the
affine-power classification in section~\ref{subsec:desitter_tails}. Formally
inserting a nonzero total coefficient into the null-contracted semiclassical
Einstein equation identifies the associated radial null curvature. Backreaction
becomes essential once the stress is large.

Our main conclusion is that cancellation of the leading term at the event
horizon does not imply cancellation at the Cauchy horizon.
Eliminating the leading pure \(V_-^{-2}\) inner-horizon divergence requires
cancellation of the total late-time flux. The fixed-background result therefore
identifies when the inner-horizon flux is amplified and when the test-field
approximation fails. Beyond that point the semiclassical double-null evolution
determines the marginal surfaces, the fate of the Cauchy horizon, and any
resulting curvature singularity, as discussed in
section~\ref{subsec:dynamical_horizons}. Questions of strong cosmic censorship
then depend on that dynamical evolution.

\acknowledgments
It is a pleasure to thank Paul Davies, Subhodeep Sarkar, Marija Tomasevic and
Tanmay Vachaspati for useful correspondence. This work is supported by the U.S.
Department of Energy, Office of High Energy Physics, under Award Number
DE-SC0019470.

\section*{Data availability}

No new data were created or analysed in this study.

\appendix

\section{\texorpdfstring{Dilaton-dependent \(s\)-wave matter terms}{Dilaton-dependent s-wave matter terms}}
\label{app:dilaton_swave}

This appendix discusses the near-horizon scaling of the simplest
dilaton-dependent terms that appear in a spherical reduction of
four-dimensional scalar matter. The
4D \(s\)-wave matter action
has the schematic 2D form
\begin{equation}
S_f
=
-\frac12
\int d^2x\,\sqrt{-g}\,
e^{-2\Phi}(\nabla f)^2,
\qquad
e^{-2\Phi}=r^2 .
\label{eq:swave_matter_action}
\end{equation}
Thus
\begin{equation}
\Phi=-\ln r .
\label{eq:Phi_def}
\end{equation}
The corresponding anomaly-induced stress tensor differs from the minimal
Polyakov stress tensor by additional local terms involving \(\Phi\) and its
derivatives
\cite{Kummer:1998sp,Fabbri:2003vy,Grumiller:2002nm,Easson:2026vte}. The local
derivative structures evaluated below do not shift the pure Polyakov geometric
offset in the constant Eddington--Finkelstein horizon coefficient; finite
nonlocal and homogeneous contributions are included separately in
\(F_{\rm other}^{(\infty)}\).

For the static double-null metric \eqref{eq:static_double_null_metric_new}, the
radial derivatives are given by \eqref{eq:partial_r_uv}. Therefore
\begin{equation}
\partial_v\Phi
=
-\frac{1}{r}\partial_v r
=
-\frac{\xi}{2r},
\qquad
\partial_u\Phi
=
-\frac{1}{r}\partial_u r
=
\frac{\xi}{2r}.
\label{eq:app_Phi_first}
\end{equation}
Both first derivatives vanish linearly in \(\xi\) at a simple horizon. A
second \(v\)-derivative gives
\begin{align}
\partial_v^2\Phi
&=
\partial_v\left(-\frac{\xi}{2r}\right)
=
\frac{\xi}{2}\partial_r\left(-\frac{\xi}{2r}\right)
\nonumber\\
&=
-\frac{\xi\xi'}{4r}
+\frac{\xi^2}{4r^2}
=
O(\xi),
\label{eq:app_Phi_second_v}
\end{align}
and similarly
\begin{equation}
\partial_u^2\Phi
=
-\frac{\xi\xi'}{4r}
+\frac{\xi^2}{4r^2}
=
O(\xi).
\label{eq:app_Phi_second_u}
\end{equation}
Mixed derivatives have the same near-horizon suppression.

Consequently, terms built from the displayed
\(\partial_v\Phi\), \(\partial_v^2\Phi\), and their products with the finite
quantity \(\partial_v\rho\) are at most \(O(\xi)\) in the
Eddington--Finkelstein component \(T_{vv}\). Writing these displayed local
derivative terms as ``loc'', they do not shift the pure Polyakov geometric
offset in the constant horizon limit,
\begin{equation}
\lim_{r\to r_h}\langle T_{vv}\rangle_{\rm P+loc}
=
t_v-\frac{N}{48\pi}\kappa_h^2
\label{eq:app_horizon_limit_local}
\end{equation}
within the Polyakov contribution plus the displayed local-derivative terms.
Finite nonlocal terms, homogeneous solutions, further state-dependent terms,
and four-dimensional scattering contributions belong in
\(F_{\rm other}^{(\infty)}\).

After transforming to \(T_{V_-V_-}\), terms that vanish as \(O(\xi)\) in
\(T_{vv}\) can still produce subleading behavior. Along the right future
Cauchy horizon, at fixed nonzero conjugate regular null coordinate, \(\xi\) is
proportional to \(V_-\). Hence an \(O(\xi)\) term in \(T_{vv}\) gives an
\(O(V_-^{-1})\) term in \(T_{V_-V_-}\). Such terms affect subleading
behavior and can obstruct full regularity, but they do not shift the pure
quadratic coefficient \(C_-\).

\section{Polyakov model and conventions}
\label{app:polyakov_summary}

This appendix summarizes the two-dimensional Polyakov model
used in the main text. For \(N\) conformal massless scalar degrees of freedom,
the trace anomaly in our curvature conventions is
\(\langle T^a{}_a\rangle=N R^{(2)}/(24\pi)\). Integrating the anomaly gives the
nonlocal functional \eqref{eq:polyakov_action}, up to Weyl-invariant and
state-dependent information. The inverse d'Alembertian requires a Green
function; equivalently, that information appears in the homogeneous part when
the action is localized \cite{Davies:1976ei,Christensen:1977jc,Callan:1992rs}.

Introducing the auxiliary field \(\psi\) gives the local action
\eqref{eq:polyakov_local}. Its field equation is \(\Box\psi=R^{(2)}\), and
metric variation gives
\begin{equation}
\langle T_{ab}\rangle_{\rm P}
=
\frac{N}{48\pi}
\left[
\nabla_a\psi\nabla_b\psi
-\frac12 g_{ab}(\nabla\psi)^2
+2g_{ab}\Box\psi
-2\nabla_a\nabla_b\psi
\right].
\label{eq:polyakov_covariant_stress}
\end{equation}
Taking the trace and using the auxiliary equation reproduces
\eqref{eq:trace_anomaly}; \(\psi\) therefore localizes the nonlocal response and
is not an additional physical matter field.

In conformal gauge \eqref{eq:conformal_metric_new}, the general solution of the
auxiliary equation is a particular solution fixed by \(\rho\) plus independent
homogeneous functions of \(u\) and \(v\). Substitution into
\eqref{eq:polyakov_covariant_stress} yields the chiral components
\eqref{eq:polyakov_Tvv_new} and \eqref{eq:polyakov_Tuu_new}, together with
\begin{equation}
\langle T_{uv}\rangle_{\rm P}
=
-\frac{N}{12\pi}\,\partial_u\partial_v\rho.
\label{eq:polyakov_Tuv_summary}
\end{equation}
The functions \(t_u(u)\) and \(t_v(v)\) are the remaining homogeneous chiral
functions. Conservation plus the anomaly fixes the geometric terms but not
these functions; boundary conditions and the quantum state do. In the
stationary family studied in this paper they are constants. Their
normal-ordered representatives obey the Schwarzian law
\eqref{eq:schwarzian_law_new}, whereas an already-renormalized RSET transforms
tensorially as emphasized in section~\ref{sec:affine}.

The equations above specify the minimal Polyakov model used in the coefficient
calculation. A spherical reduction of four-dimensional matter is generally
dilaton coupled, and a full four-dimensional RSET contains global mode and
scattering information. These additional contributions are collected in
\(F_{\rm other}^{(\infty)}\) in section~\ref{sec:totalflux}.

\bibliographystyle{JHEP}
\bibliography{2dbhs}

\end{document}